\documentclass[12pt]{article}
\usepackage{amsmath,amsfonts,amsthm,amssymb,slashed,graphicx}
\usepackage{xcolor}
\usepackage{graphicx}
\usepackage{epstopdf}
\usepackage{array,booktabs}
\usepackage{hyperref}
\usepackage{tabulary}
\usepackage{multirow}
\usepackage{array,arydshln}
\usepackage{bbm} 
\usepackage{cite}
\usepackage{mathtools} 

\hypersetup{linktoc = all}  
\hypersetup{pdfborderstyle={/S/U/W 0.5}}

\numberwithin{equation}{section}
\def\a{\alpha}

\newcommand{\beq}{\begin{equation}}
\newcommand{\eeq}{\end{equation}}
\newcommand{\bea}{\begin{eqnarray}}
\newcommand{\eea}{\end{eqnarray}}

\newcommand{\nn}{\nonumber}

\newcommand{\address}[1]{\vbox{\center\em#1}}
\renewcommand{\title}[1]{\vbox{\center\LARGE{#1}}\vspace{5mm}}

\DeclareMathOperator{\Tr}{Tr}

\newcommand{\rf}[1]{\eqref{#1}}
\declareslashed{}{/}{-.08}{0}{V} 

\makeatletter

\newcommand*{\letterdef@}{}
\newcommand*{\letterdef}[3]{%
  \def\letterdef@##1{\expandafter\newcommand\csname #1\endcsname{#2{##1}}}%
  \@tfor\@tempa :=#3\do{\expandafter\letterdef@\expandafter{\@tempa}}}
\makeatother
\letterdef{b#1} {\mathbb} {CHPRZ}    
\letterdef{c#1} {\mathcal}{ABCDEFGHIJKLMNOPQRSTUVWXYZ} 
\letterdef{rm#1}{\mathrm} {dDmM} 
\letterdef{#1frak}{\mathfrak}{t}

\begin{document}
\bibliographystyle{utphys}

\begin{titlepage}
\begin{center}
\vspace{5mm}
\hfill {\tt }\\
\vspace{8mm}

\title{On ${\cal{N}}=2$ supersymmetric gauge theories on $S^2\times S^2$}
\vspace{6mm}

Musema
Sinamuli${}$\footnote{\href{mailto:cmusema@perimeterinstitute.ca}
{\tt cmusema@perimeterinstitute.ca}}
\vskip 15mm
\address{
${}$Perimeter Institute for Theoretical Physics,\\
Waterloo, Ontario, N2L 2Y5, Canada}

\vskip 10mm

\address{${}$Department of Physics, University of Waterloo,\\
Waterloo, Ontario N2L 3G1, Canada}

\end{center}

\vspace{5mm}
\abstract{ \normalsize{We construct a supergeometry based on $S^2\times S^2$ on which four dimensional ${\cal N}=2$ gauge theories can be placed supersymmetrically while preserving all supersymmetries.   By embedding the supergeometry in four dimensional ${\cal N}=2$ supergravity we are able to construct an arbitrary $\cN=2$ gauge theory on $S^2\times S^2$.  We show that $\cN=2$ gauge theories are invariant under the exceptional superalgebra $D(2,1,\alpha)$, where $\alpha$ is the ratio of the radii of the two $S^2$'s.   We solve the supersymmetry fixed points equations for a choice of supercharge in $D(2,1,\alpha)$. The solution of these BPS equations, which we find,  would serve as the exact saddle point configurations of a    localization computation of the partition function of $\cN=2$ gauge theories on $S^2\times S^2$. }
\noindent
}
\vfill
 
\end{titlepage}

\section{Introduction}

The computation of the partition function of four dimensional $\cN=2$ gauge theories on $S^4$ by Pestun \cite{Pestun:2007rz} has led to new insights into the non-perturbative dynamics of gauge theories. 
A natural avenue of investigation is to consider more general curved backgrounds over which a four dimensional $\cN=2$ gauge theory can be placed, and to compute the corresponding exact partition function. Just as the $S^4$ partitition function of $\cN=2$ superconformal field theories computes the exact K\"ahler potential on the conformal manifold \cite{Gerchkovitz:2014gta}\cite{Gomis:2014woa}, it is interesting to understand the intrinsic physical  meaning of the partition function of such theories on other backgrounds. 

In this paper we identify a supergeometry based on $S^2\times S^2$  over which an arbitrary four dimensional $\cN=2$ gauge theory   can be placed while preserving all supercharges. The theory is constructed by embedding our $S^2\times S^2$ background in four dimensional $\cN=2$ supergravity (supersymmetric backgrounds in  $\cN=2$ supergravity have been considered in~\cite{Hama:2012bg}\cite{Klare:2013dka}). We show that the theory is   invariant under the exceptional superalgebra  $D(2,1,\alpha)$, where $\alpha$ is the ratio of the radii of the two $S^2$'s. We solve the supersymmetric fixed point equations for a choice of supercharge in $D(2,1,\alpha)$. We find that the non-singular field configurations are labeled by quantized magnetic flux over each of the two $S^2$'s and have no continuous moduli. We also show that point-like instanton and anti-instanton configurations are supersymmetric at poles on $S^2\times S^2$.

The plan of the rest of the paper is as follows. In section 2 we construct a supergeometry on $S^2\times S^2$ by specifying Killing spinor equations and identify the supergeometry with a coset superspace. In section 3 we embed the Killing spinor equations in section 2 in four dimensional $\cN=2$ supergravity. This requires finding the off-shell field configurations for the supergravity multiplet. We also identify $D(2,1,\alpha)$ as the supersymmetry algebra of four dimensional $\cN=2$ gauge theories on $S^2\times S^2$. In section 4 we write down the supersymmetry transformations of the matter multiplets on this background. In section 5 we solve the supersymmetry fixed point equations for a choice of supercharge in $D(2,1,\alpha)$, which would serve as the exact saddle points in the localization computation of $\cN=2$ gauge theories on $S^2\times S^2$.  Various computational details are relegated to the Appendices
 
\noindent
  Recently a paper studying the topologically twisted theory on $S^2\times S^2$ has appeared in \cite{Bawane:2014uka}.
  
\section{Supergeometry on $S^2\times S^2$}

Supersymmetry transformations in ${\cal N}=2$ theories are parametrized by Killing spinors $\epsilon^i$ and $\epsilon_i$ 
of opposite chirality 
\beq
\gamma_* \epsilon^i=\epsilon^i\qquad \gamma_* \epsilon_i=-\epsilon_i
\eeq
transforming as doublets of the  $SU(2)_R$ R-symmetry (see Appendix \ref{conventions} for conventions and notations). 

On the background geometry $S^2\times S^2$ of radii $\tilde r$ and $r$ we   define the following consistent set of Killing spinor equations 
\begin{align}
\label {killing}
\nabla_{m}\epsilon ^{i}&=\frac{1}{2 \tilde{r}}\Gamma\gamma_{m} {\epsilon}^{ij}\epsilon_j\,,\qquad\qquad  \nabla_{m}\epsilon_{i}=\frac{1}{2 \tilde{r}}\Gamma\gamma_{m} {\epsilon}_{ij}\epsilon^j\qquad\qquad m=0,1 \nn\\
\nabla_{p}\epsilon ^{i}&=\frac{i}{2 r}\Gamma\gamma_{p}  {\epsilon}^{ij}\epsilon_j\,,\qquad\qquad  \, \nabla_{p}\epsilon_{i}=\frac{i}{2 r}\Gamma\gamma_{p}  {\epsilon}_{ij}\epsilon^j\qquad\qquad\ \  p=2,3
\end{align}
 where $\Gamma\equiv\gamma_{\hat 0}\gamma_{\hat 1}$.  These equations can be diagonalized by combining the Killing spinors into $SU(2)_R$ doublets
\beq
\chi^i= \epsilon^i+ i {\epsilon}^{ij} \epsilon_j \,,
\label{doubb}
\eeq
such that
\beq
\epsilon^i=\chi^i_L\qquad\qquad \epsilon_i=i \epsilon_{ij} \chi^j_R\,.
\eeq 
These obey
\begin{align}
\label {killingdiaga}
\nabla_{m}\chi^i&=-   \frac{i}{2 \tilde{r}}\gamma_{m} \Gamma\chi^i \qquad\qquad m=0,1 \\
\nabla_{p}\chi^i&= \frac{1}{2 r}\gamma_{p} \Gamma \chi^i \qquad\qquad \ \ \ p=2,3\nn\,.
\end{align}
By choosing the  following basis of $\gamma$-matrices
\begin{align}
 \gamma_{\hat 0}&= -\tau_2\otimes 1\\
 \gamma_{\hat 1}&= \tau_1\otimes 1\\
 \gamma_{\hat 2}&= \tau_3\otimes \tau_1\\
 \gamma_{\hat 3}&= \tau_3\otimes \tau_2\\
 \gamma_*&=- \gamma_{\hat 0}\gamma_{\hat 1} \gamma_{\hat 2} \gamma_{\hat 3}=\tau_3\otimes \tau_3\,,
\end{align}
 the Killing spinor equations \rf{killing} decouple between the two $S^2$'s 
\begin{align}
\nabla_{\hat 0}\chi^i&= {i\over 2\tilde{r}}(\tau_1\otimes 1)\chi^i\\
\nabla_{\hat 1}\chi^i&= {i\over 2\tilde{r}}(\tau_2\otimes 1)\chi^i\\
\nabla_{\hat 2}\chi^i&= {i\over 2r}(1 \otimes \tau_1)\chi^i\\
\nabla_{\hat 3}\chi^i&= {i\over 2r}(1 \otimes \tau_2)\chi^i\,.
\end{align}
In the vielbien  frame 
 \beq
 e^{\hat 0}=\tilde r d\tilde \theta\qquad e^{\hat 1}=\tilde r \sin\tilde \theta d\tilde \phi\qquad  e^{\hat 2}= r d  \theta\qquad e^{\hat 3}=  r \sin \theta d  \phi\,
 \eeq
 we have that\footnote{Killing spinors on $S^2$ can be found in \cite{Benini:2012ui,Doroud:2012xw}.}
 \beq
 \chi^i=e^{{i \over 2} \tau_1\tilde\theta}e^{{i \over 2} \tau_3\tilde\phi}\otimes e^{ {i \over 2} \tau_1\theta}e^{{i \over 2}  \tau_3\phi} \chi_{(0)}^i\, 
  \label{killsp}
 \eeq
where $ \chi_{(0)}^i$ is a constant $SU(2)_R$ doublet of Dirac spinors. In Appendix \ref{stereographic} we express the spinors in the stereographic coordinate system and show that the spinors are non-singular everywhere on $S^2\times S^2$.

The Killing spinors we have constructed are acted on by the $SU(2)_1\times SU(2)_2$ isometries   of $S^2\times S^2$ through  the Lie-Lorentz derivative. For a Killing vector field $\xi$ this derivative acts by
\beq
{\cal L}_\xi =\nabla_\xi+{1\over 4} \nabla_\mu \xi_\nu \gamma^{\mu\nu}\,.
\eeq
This analysis implies that the Killing spinors transform in the $(2,2,2)$ representation of  $SU(2)_1\times SU(2)_2\times SU(2)_R$.
 The  associated supergeometry is the coset superspace
 \beq
 {D(2,1,\alpha)\over U(1)\times U(1)}\,,
 \eeq
where $\alpha={\tilde r\over r}$.
   
  \section{${\cal N}=2$ Supergravity Background Fields for $S^2\times S^2$}
  
 ${\cal N}=2$ gauge theories on $S^2\times S^2$ are based on a vectormultiplet and a hypermultiplet.
    Supersymmetry requires non-minimal couplings of  the vectormultiplet and hypermultiplet to the background geometry. These can be found by the Noether procedure starting from the supersymmetry transformations and action of the theory in flat space. 
  
 A less laborious and more conceptual way of proceeding    is to embed the supergeometry we have just constructed as a supersymmetric background of off-shell ${\cal N}=2$ supergravity, in the spirit advocated in \cite{Festuccia:2011ws}. This approach relies on the  already known supersymmetry transformations and couplings of 
 a vectormultiplet and hypermultiplet to an off-shell supergravity multiplet. For our construction we   consider the coupling of a vectormultiplet and hypermultiplet  to the ${\cal N}=2$ Weyl multiplet~\cite{deWit:1980tn} (we refer to \cite{freedman12} for  more details).
 
 Off-shell ${\cal N}=2$ superconformal transformations are realized on the Weyl multiplet, whose independent fields are
	\begin{align}
	\text{bosonic:} &\, e^a_\mu, b_\mu,V_{\mu\, i}^{\ \ j}, A^R_\mu, T_{ab},\bar T_{ab}, D\cr
	\text{fermionic:} &\, \psi_\mu^i, \psi_{\mu\,i}, \chi^i,\chi_i\,.
	\label{weylm}
	\end{align}
The fields $e^a_\mu, b_\mu,V_{\mu\, i}^{\ \ j}, A^R_\mu, \psi_\mu^i, \psi_{\mu\,i}$ are the gauge fields for translations, dilatations, $SU(2)_R$, $U(1)_R$ and Poincar\'e supersymmetry generators in the ${\cal N}=2$  superconformal algebra. The Weyl multiplet also includes  the bosonic auxiliary fields $T_{ab}, \bar T_{ab}$ and  $D$, and the fermionic auxiliary fields $\chi^i$ and $\chi_i$. In Euclidean signature  $T_{ab}$ is a self-dual  and $\bar T_{ab}$ is an anti-self-dual rank-two tensor.  

Supersymmetric (bosonic) backgrounds are   background values of the Weyl multiplet obeying
	 \beq
	 \left(\delta_\epsilon+\delta_\eta\right) \psi^i_\mu=0\qquad \left(\delta_\epsilon+\delta_\eta\right) \chi^i=0\,,	
	  \eeq
	where    $(\epsilon^i,\epsilon_i)$ and $(\eta^i,\eta_i)$ parametrize the Poincar\'e and conformal supersymmetry transformations.\footnote{The supersymmetry transformations are the same as those in Lorentzian signature, but now $(\epsilon_i,\epsilon^i)$  and $(\eta^i,\eta_i)$ are not related by conjugation, and are independent spinors.
Also, we allow all fields (except the metric) to be complex. In Euclidean signature, however, $T_{ab}$ is selfdual while
$\bar T_{ab}$ is anti-selfdual, and are  independent fields (see Appendix \ref{conventions}).}
 	The explicit form of these transformations are~\cite{deWit:1979ug}\cite{deWit:1979pq}\cite{deWit:1982na} (we use \cite{parisLect})
\begin{align}
 \delta \psi_\mu^i=& \left(\partial_\mu+{1\over 4}\gamma^{ab} w_{\mu\, ab} -i{1\over 2}A^R_\mu +{1\over 2} b_\mu  \right)\epsilon^i+V^i_{\mu  \ j} \epsilon^j- {1\over 16} \gamma^{ab}T_{ab} \epsilon^{ij} \gamma_\mu \epsilon_j -\gamma_\mu \eta^i\,, \nn\\
 \delta \chi^i =& \frac{1}{2} D \epsilon^i + \frac{1}{6} \gamma^{ab} \left[ -\frac{1}{4} \slashed {\cal D}   T_{ab}^- \epsilon^{ij}\epsilon_j
			- \widehat{R}_{ab} (U_j^{\;\,i})\epsilon^j+ i \widehat{R}_{ab}(T) \epsilon^i + \frac{1}{2} T_{ab}^- \epsilon^{ij} \eta_j\right]	 \,,			\label{BPSsugra}
\end{align}
where  $\cal D$ is the superconformal covariant derivative and $\widehat{R}_{ab}(T)$ and $ \widehat{R}_{ab} (U_j^{\;\,i})$ are covariant curvatures for   $U(1)_R$ and $SU(2)_R$. 
 
Our goal is to embed the  Killing spinor equations on $S^2\times S^2$ in \rf{killing}
as a supersymmetric background for the Weyl multiplet. By analyzing \rf{BPSsugra} we find indeed that the following background fields give rise to our  supersymmetric $S^2\times S^2$ supergeometry
\beq
e^a_m=e^a_m|_{S^2\times S^2}\,,\ T_{\hat{0} \hat{1}}=T_{\hat 2\hat 3}=\left(\frac{i}{r}+\frac{1}{\tilde{r}}\right)\,,\   \bar T_{\hat{0} \hat{1}}=-\bar T_{\hat{2} \hat{3}}=\left(\frac{i}{r}+\frac{1}{\tilde{r}}\right)\,,\ D =\frac{1}{6}\left( {1\over r^2}+{1\over \tilde r^2}\right)\,.
\label{background}
\eeq
From \rf{BPSsugra} we find that the conformal supersymmetry parameters   that give rise to $S^2\times S^2$  are 
\beq
\eta^i=\frac{1}{4}\left(\frac{i}{r}-\frac{1}{\tilde{r}}\right)\Gamma \epsilon^{i j}\epsilon_{j}\,, \qquad \eta_i=\frac{1}{4}\left(\frac{i}{r}-\frac{1}{\tilde{r}}\right)\Gamma \epsilon_{i j}\epsilon^{j}\,.
\label{backgrounded}
\eeq

The supergravity approach also provides us with a systematic way of identifying the superisometry algebra of a supersymmetric background of supergravity.
The structure constants of the ${\cal N}=2$ superconformal transformations  generated by the closure of the supergravity transformations  determine the rigid supersymmetry algebra of our $S^2\times S^2$ background. This is obtained by  evaluating the structure constants on the $S^2\times S^2$ background fields \rf{background}\rf{backgrounded}. The supergravity commutators yield\footnote{We omit special conformal transformations, as they act trivially on vectormultiplet and hypermultiplet fields. Actually, only the dilatation gauge field $b_\mu$ is acted on by special conformal transformations.}
\beq
[\delta_1,\delta_2]= \xi^m P_m+\lambda_a R^a+\lambda _{{\rm D}} D+\lambda _{R} R+ {1\over 2}\lambda^{ab} L_{ab}\,,
\label{susyalg}
\eeq
where $\delta\equiv\delta_\epsilon+\delta_\eta$. The parameters $(\xi^m,\lambda_a,\lambda _{{\rm D}},\lambda^{ab})$ are completely determined by the  off-shell  supergravity transformations of the Weyl multiplet.

Using the $S^2\times S^2$ Killing spinor equations \rf{killing},  it follows  that the vector field
produced by two superconformal transformations
\beq
\xi_m={1\over 2} \bar\epsilon_2^i \gamma_m \epsilon_{1i}+{1\over 2}\bar\epsilon_{2 i} \gamma_m \epsilon_1^i
\label{killv}
\eeq
is a Killing vector on $S^2\times S^2$.  Using the Killing spinors   \rf{killsp} we compute $\xi_m$ in  
Appendix~\ref{killingvect}.

Evaluating the parameters  in \rf{susyalg} on the $S^2\times S^2$ background we find that    dilatations  $D$ and the  $U(1)_R$ R-symmetry are broken, while    the $SU(2)_R$ R-symmetry is  unbroken 
\begin{align}
\lambda _{{\rm D}}=& - {1\over 2}\left( \bar \epsilon ^i_1\eta _{2i}+\bar \epsilon _{1i}\eta _2^i-    \bar \epsilon ^i_2 \eta _{1i}
-\bar \epsilon_{2i} \eta ^i_1\right)=0\,,\cr
\lambda _{R} =&  {i\over 2}\left( \bar \epsilon ^i_1\eta _{2i}-\bar \epsilon _{1i}\eta _2^i -\bar \epsilon ^i_2\eta _{1i}+\bar \epsilon _{2i}\eta _1^i\right)=0\,,\cr
  \lambda _{j}{}^i =&  -\bar \epsilon ^i_1\eta _{2j}+\bar \epsilon _{1j}\eta _2^i+\bar \epsilon ^i_2\eta _{1j}-\bar \epsilon _{2j}\eta _1^i =\frac{1}{2}\left(\frac{i}{r}-\frac{1}{\tilde{r}}\right)\left(-\varepsilon _{jk}\bar \epsilon ^{(i}_1\Gamma \epsilon _2^{k)}
                                    +\bar \epsilon _{1(j}\Gamma \epsilon _{2k)} \varepsilon^{ik}\right)\,.
 \end{align}
 Therefore,  we have shown starting from supergravity that the  rigid supersymmetry algebra on our $S^2\times S^2$ is the complexified $D(2,1,\alpha)$ supersymmetry algebra, with $\alpha={\tilde r\over r}$. The eight conserved   supercharges, which transform in the $(2,2,2)$ representation of $SU(2)_1\times SU(2)_2\times SU(2)_R $ close into  the  $SU(2)_1\times SU(2)_2$ isometries of $S^2\times S^2$ and the $SU(2)_R$ R-symmetry.\footnote{On the fields, a   local Lorentz transformation   is also induced.}

\section{${\cal N}=2$ Gauge Theories on $S^2\times S^2$}

Embedding $S^2\times S^2$ as a supersymmetric  background in ${\cal N}=2$ supergravity allows us to immediately write down  the   $D(2,1,\alpha)$ supersymmetry transformations acting on the vectormultiplet  and  hypermultiplet fields. These can be obtained from the supergravity literature by plugging in the $S^2\times S^2$ background fields 
\rf{background}\rf{backgrounded} on the known superconformal supergravity transformations. 

An off-shell ${\cal N}=2$ vectormultiplet consists of
\begin{align}
	\text{bosonic:} &\, X, A_\mu, Y_{ij} \cr
	\text{fermionic:} &\, \Omega_i
	\label{vector}
	\end{align}
 a complex scalar $X$, a gauge field $A_\mu$, a triplet of real auxiliary fields $Y_{ij}=Y_{ji}$ and gauginos $\Omega_i$.  All fields  in the multiplet transform in the adjoint representation of a gauge group $G$.  An on-shell ${\cal N}=2$ hypermultiplet consists of
\begin{align}
	\text{bosonic:} &\, q_i\cr
	\text{fermionic:} &\, \psi
	\label{vector}
	\end{align}
 a doublet of   scalars $q_i $  and hyperinos $\psi$.  The hypermultiplet can be coupled to a   vectormultiplet by embedding the gauge group $G$ in the symplectic symmetry group acting on the hypermultiplets.   Fields in the hypermultiplet transform in a representation $R$ of $G$. We consider the   vectormultiplet  and  a hypermultiplet  coupled to   the Weyl multiplet.

On a vectormultiplet  and hypermultiplet  the commutator of two supergravity transformations yields \rf{susyalg} {\it together} with a gauge transformation acting in the appropriate representation of the gauge group $G$. The induced field dependent gauge transformation parameter is
\beq
\label{gaugepar}
\Lambda= X \epsilon^{ij} \bar \epsilon_{2i} \epsilon_{1j}+ \bar X \epsilon_{ij} \bar \epsilon_{2}^i \epsilon_{1}^j\,,
 \eeq
 which when written in terms of the doublets \rf{doubb} is 
 \beq
\Lambda={1\over 2} (\bar X - X) \epsilon_{ij}\bar \chi_2^i\chi_1^j+{1\over 2} (X+\bar X) \epsilon_{ij}\bar \chi_2^i\gamma_*\chi_1^j\,.
\eeq
This gauge transformation   plays an important role in the computation of the partition function of $\cN=2$ gauge theories on $S^2\times S^2$.

The supersymmetry transformations and invariant action for the vectormultiplet and hypermultiplet can be obtained from Chapter 20 of \cite{freedman12}. For future reference,    the $D(2,1,\alpha)$ supersymmetry transformations acting on the  gauginos in the vectormultiplet are 
\begin{align}
 \delta \Omega_i &={1\over 4} \left[ F_{ab}-{1\over 2} \bar X T_{ab} \right]\gamma^{ab}\epsilon_{ij}\epsilon^j+\gamma^\mu D_\mu X\epsilon_i-i [X,\bar X] \epsilon_{ij} \epsilon^j+Y_{ij} \epsilon^j +{2X}\eta_i\\
  \delta \Omega^i &={1\over 4} \left[ F_{ab} -{1\over 2}   X \bar T_{ab}\right]\gamma^{ab}\epsilon^{ij}\epsilon_j+\gamma^\mu D_\mu \bar X\epsilon^i+i[X,\bar X] \epsilon^{ij} \epsilon_j+Y^{ij} \epsilon_j  +2\bar X \eta^i\,.
 \end{align}
  $(\epsilon^i, \epsilon_i)$ are the Killing spinors on $S^2\times S^2$ we constructed, $(\eta^i,\eta_i)$ are given in \rf{backgrounded} and   $T_{ab}$ and $\bar T_{ab}$   in~\rf{background}.

The action of the vectormultiplets and action and on-shell supersymmetry transformations for the hypermultiplet are given in chapters 20.2.4 and 20.2.3 of 
\cite{freedman12} by substituting the background fields \rf{background} on $S^2\times S^2$. 
 
\section{Supersymmetric Fixed Points on $S^2\times S^2$}
  
  In this final section we find the supersymmetric field configurations associated to a particular supersymmetry transformation in $D(2,1,\alpha)$. These field configurations correspond  to the exact saddle points of the partition function of ${\cal N}=2$ gauge theories on $S^2\times S^2$ when computed by supersymmetric localization with the corresponding supercharge.
  
  The supersymmetry transformation that we consider is generated by the following choice of constant spinor $\chi_{(0)}^j$ in 
  \rf{killsp}
  \beq
  \chi_{(0)}^j=\delta^j_1\left[{1+\tau_3\over 2}\otimes   {1+\tau_3\over 2}  \right] \chi_A^j+\delta^j_2\left[{1-\tau_3\over 2}\otimes   {1-\tau_3\over 2}  \right] \chi_B^j\,,
  \label{zeromode}
  \eeq
  which projects onto the sum   $(\uparrow,\uparrow,\uparrow)\oplus (\downarrow,\downarrow,\downarrow)$ under $SU(2)_1\times SU(2)_2\times SU(2)_R$. The corresponding transformation obeys
  \beq
  \delta^2= \tilde J_3+J_3+R\,,
  \eeq
  where $\tilde J_3$ and $J_3$  are the Cartan generators of the $SU(2)$ isometry acting on $S^2_{\tilde r}$ and $S^2_{r}$ respectively, while $R$ is the $SU(2)_R$ R-symmetry Cartan generator. The equivariant parameters induced by the supersymmetry transformation  for $\tilde J_3$ and $J_3$  are given by
  \beq
  \varepsilon_1={i\over \tilde r}\qquad \varepsilon_2={1\over \tilde r}
  \label{equivariant}
  \eeq
  while the equivariant parameter for $SU(2)_R$ is 
  \beq
  \varepsilon_1+\varepsilon_2\,.
  \eeq
  On the vectormultiplet and hypermultiplet the induced gauge transformation is 
   \beq
\Lambda= (\bar X - X) \cos\theta+(X+\bar X) \cos\tilde\theta\,.
\label{gauget}
 \eeq
 
 The Killing spinor corresponding to the choice \rf{zeromode} is non-chiral everywhere on   $S^2\times S^2$ except at the four fixed points of $\delta^2$, labeled by a North and/or South pole for each of the $S^2$'s. At these four fixed points the non-vanishing Killing spinor  has  a definite  chirality:
 \beq
 \text{NN}: L\qquad  \text{NS}: R\qquad  \text{SN}: R\qquad  \text{SS}: L\,.
 \eeq
  Since chiral (anti-chiral) supersymmetry transformations  correspond to instanton (anti-instanton) field configurations, this implies that on $S^2\times S^2$ there are supersymmetric point-like instanton configurations at   \text{NN} and \text{SS} poles and 
  supersymmetric point-like anti-instanton configurations at the   \text{NS} and \text{SN} poles.
  
  We now analyse the supersymmetric fixed point equations for our choice of supersymmetry transformation \rf{zeromode}. We have already identified singular instanton and anti-instanton field configurations at the poles of $S^2\times S^2$, so we now turn to the analysis of the smooth supersymmetric field configurations. This requires solving the equations 
  \beq
  \delta \Omega^i=\delta \Omega_i=0
  \eeq
  for our   choice of transformation. We write the supersymmetry equations in Appendix \ref{BPSequ}. The 
  most general smooth solution is labeled by a pair of vector of integers $(\tilde B, B)$ that represent quantized flux over each of the two $S^2$'s. These fluxed  take values in the Cartan subalgebra of the gauge group $G$.
 The supersymmetry equations also fix the scalar field in the vectormultiplet in terms of the fluxes. Explicitly\footnote{This is similar to the Coulomb branch localization saddle points of two dimensional $\cN=(2,2)$ gauge theories on  $S^2$ \cite{Benini:2012ui,Doroud:2012xw}.}
  \beq
     F_{\hat 0\hat 1}={2\over \tilde r} \text{Re}(X)={\tilde B \over 2\tilde r^2} \qquad  F_{\hat 2\hat 3}={2\over  r} \text{Im}(X)={B\over 2 r^2}
     \label{susyconfg}
  \eeq
  In solving the supersymmetry equations we have used the standard reality properties on the fields, in particular $X^\dagger=\bar X$.
  As in the case of $S^4$ \cite{Pestun:2007rz} and squashed $S^4$\cite{Hama:2012bg}, there are no non-trivial solutions to the supersymmetry equations for the fields in the hypermultiplet.

  Our analysis suggests the following answer for the partition of ${\cal N}=2$ gauge theories on $S^2\times S^2$. It is given by the sum over all quantized fluxes $B$ and $\tilde B$ of the product the  instanton partition function at the NN and SS poles with the   anti-instanton partition function at the NS and SN poles. These Omega-background instanton partition functions \cite{Nekrasov:2002qd} are to evaluated for the values of the equivariant parameters induced by the $S^2\times S^2$ geometry. The geometrical ones are governed by \rf{equivariant}. The equivariant gauge transformation is obtained by evaluating \rf{gauget} on the supersymmetric field configurations \rf{susyconfg}.
 When evaluated on the poles, we get
 \beq
 \Lambda_{NN}=-\Lambda_{SS}=2\bar X~~~~~~~\Lambda_{NS}=-\Lambda_{SN}=2X
 \eeq
 and therefore the equivariant gauge parameters are quantised. It would be interesting to confirm this intuition by a detailed
 supersymmetric localization computation

\section{Acknowledgements}
We would like to thank Jaume Gomis   and Antoine Van Proeyen for discussions. We are grateful to The Perimeter Institute for Theoretical Physics for supporting the work and providing us with the work environment.   This research was supported in part by Perimeter Institute for Theoretical Physics. Research at Perimeter Institute is supported 
by the Government of Canada through Industry Canada and by the Province of Ontario through the Ministry of Research and Innovation.

\vfill\eject

\appendix
 
\addtocontents{toc}{\protect\setcounter{tocdepth}{1}}

\section{Notations and Conventions}
\label{conventions}

Curved indices are $\mu,\nu,\ldots$ while tangent space indices are $a,b,\ldots$. We also denote tangent space indices with a  hat. e.g., $\hat 0$. We split $\mu=(m,p)$, where $m=0,1$ parametrizes $S^2_{\tilde r}$ and $p=2,3$ parametrizes $S^2_{r}$

We take the Lorentzian chirality matrix to be
\beq
\gamma_*=i \gamma_{\hat 0}\gamma_{\hat 1}\gamma_{\hat 2}\gamma_{\hat 3}=-i \gamma^{\hat 0}\gamma^{\hat 1}\gamma^{\hat 2}\gamma^{\hat 3}\,.
\eeq
We continue to Euclidean signature by changing coordinates
\beq
x^0=-ix^0_E\,,
\eeq
so that
\beq
\gamma_{\hat 0\, E}=-i\gamma_{\hat 0}\qquad \gamma^{\hat 0}_E=i\gamma^{\hat 0}
\eeq

This implies that  Euclidean chirality matrix is
\beq
\gamma_*=-\gamma_{{\hat 0}\, E}\gamma_{{\hat 1}\, E}\gamma_{{\hat 2}\, E}\gamma_{{\hat 3}\, E}=-\gamma^{{\hat 0}\, E}\gamma^{{\hat 1}\, E}\gamma^{{\hat 2}\, E}\gamma^{{\hat 3}\, E}\,.
\eeq
We drop the index $E$ to avoid cluttering. The chirality of the various fermions is
\begin{center}
  \begin{tabular}{*{16}{>{\(}c<{\)}}}
    \toprule
    \multicolumn{8}{c}{SUSY} &
    \multicolumn{6}{c}{gravity multiplet} &
    \\
    \epsilon^i &  \epsilon_i  & \eta^i & \eta_i & Q^i & Q_i & S^i & S_i &
    \psi_\mu^i  & \psi_{\mu i} & \chi^i & \chi_i      \\
   L & R  & R &L &
    R & L & L & R & L & R & L&R
     \\
    \bottomrule
  \end{tabular}
\end{center}

A $L$ and $R$ chiral fermion obeys
\beq
P_L \psi=\psi=\gamma_*\psi\qquad P_R \psi=\psi=-\gamma_*\psi\,
\eeq
where
\beq
P_L={1\over 2}(1+\gamma_*)\qquad P_R={1\over 2}(1-\gamma_*)\,.
\eeq

\noindent
{\bf Epsilon tensor}: Defined to obey $\epsilon^{12}=\epsilon_{12}=1$. It satisfies
\beq
\epsilon^{ik}\epsilon_{kj}=-\delta^i_j\,.
\eeq
\section{Killing Spinors in stereographic coordinates}
\label{stereographic}

  The metric on $S^2\times S^2$  in stereographic coordinates reads  
\begin{equation}
ds^2=4{\tilde{r}}^{2}\frac{dw d\bar{w}}{{(1+|w|^2)}^2}+4 r^{2}\frac{dz d\bar{z}}{{(1+|z|^2)}^2}\nonumber\,.
\end{equation}
The stereographic coordinates $(w,z)$ cover a patch including the north pole of each $S^2$ and are given in terms of the spherical coordinates by 
\begin{eqnarray}
w&=&\tan\frac{\tilde{\theta}}{2}~e^{i\tilde{\phi}}={1\over u}\nonumber\\
z&=&\tan\frac{\theta}{2}~e^{i\phi}={1\over v}\nonumber\,.
\end{eqnarray}
Coodinates $(u,v)$ cover a  patch including the south pole of each $S^2$.

Vielbeins   regular at the (N,N)  poles $(w=0, z=0)$ are 
\beq{\hat{e}}^{0}=\frac{2\tilde{r}}{1+\bar{w}w}dw_R ,~  {\hat{e}}^{1}=\frac{2\tilde{r}}{1+\bar{w}w}dw_I ,~ {\hat{e}}^{2}=\frac{2r}{1+\bar{z}z}dz_R ,~ {\hat{e}}^{3}=\frac{2r}{1+\bar{z}z}dz_I\,.
\eeq 
 
\vskip 5pt In terms of the original ones they are written as an $SO(2)\times SO(2)$ rotation, given by
\begin{eqnarray}
{\hat{e}}^{0}&=\cos\tilde\phi ~e^{0}-\sin\tilde\phi ~e^{1}\qquad{\hat{e}}^{2}&=\cos\phi ~ e^{2}-\sin\phi ~ e^{3}\  \nonumber\\
{\hat{e}}^{1}&=\sin\tilde\phi ~e^{0}+\cos\tilde\phi ~e^{1}\qquad {\hat{e}}^{3}&=\sin\phi ~ e^{2}+\cos\phi ~ e^{3}\nonumber\,.
\end{eqnarray}
 
  Under such a change of frame,  the Killing spinors  transform under an $SO(2)\times SO(2)$ rotation as spinors. 
Hence 
\beq
 \hat{\chi}^i =\exp\left(-\frac{\tilde\phi}{2}\gamma_{{\hat 0} {\hat 1}}\right)\exp\left(-\frac{\phi}{2}\gamma_{{\hat 2} {\hat 3}}\right)\chi^i \nonumber
\eeq 
where
\beq
 \chi ^i=e^{ {i \over 2} \tau_1\tilde\theta}e^{{i \over 2} \tau_3\tilde\phi}\otimes e^{{i \over 2} \tau_1\theta}e^{{i \over 2}  \tau_3\phi} \chi_{0}^i\nonumber
  \eeq
 and
\begin{eqnarray}
\gamma_{\hat 0 \hat 1}&=i\tau_{3}\otimes I\nonumber\\
\gamma_{\hat 2 \hat 3}&=iI\otimes \tau_{3}\nonumber\,.
\end{eqnarray}
We need to calculate 
\begin{eqnarray}
 \ \ e^{-{i \over 2} \tau_3\phi}e^{{i \over 2} \tau_1\theta}e^{{i \over 2}  \tau_3\phi}&=\begin{pmatrix}\cos{\theta\over 2}& i \sin   {\theta\over 2} e^{-i\phi}\\ i \sin   {\theta\over 2} e^{i\phi}&\cos{\theta\over 2}\end{pmatrix}= \cos{\theta\over 2}\begin{pmatrix}1&i \bar z\\ i z& 1\end{pmatrix}\\
\end{eqnarray}
and likewise for the other $S^2$. Note that the prefactor is non-vanishing around the corresponding poles, and the 
matrix elements combine into regular functions of stereographic coordinates. This implies that the spinors around  all poles are regular. 

 For the  $(\uparrow,\uparrow,\uparrow)$ spinor we have (tensored with 
$\begin{pmatrix}1\cr 0
 \end{pmatrix}$) for the $SU(2)_R$ R-symmetry)
\begin{eqnarray}
\hat \chi^1=\frac{1}{\sqrt{(1+\bar w w)(1+\bar z z)}}\begin{pmatrix}1\cr iz\cr iw\cr -wz
 \end{pmatrix}\,,
\end{eqnarray}
while for the $(\downarrow,\downarrow,\downarrow)$ spinor we have (tensored with 
$\begin{pmatrix}0\cr 1
 \end{pmatrix}$  for the $SU(2)_R$ R-symmetry)
\begin{eqnarray}
\hat \chi^2=\frac{1}{\sqrt{(1+\bar w w)(1+\bar z z)}}\begin{pmatrix}-\bar w\bar z\cr i\bar w\cr i\bar z\cr 1
 \end{pmatrix}\,.
\end{eqnarray}
We have used that
\beq
\cos{\theta\over 2}={1\over \sqrt{1+z\bar z}}\qquad \sin{\theta\over 2}={1\over \sqrt{1+v\bar v}}\,.
\eeq

    \section{Killing Vectors from Killing Spinors}
    \label{killingvect}
 
The Killing vector obtained by two SUSY transformations is (using that $\bar\epsilon \gamma^\a \lambda=-\bar\lambda \gamma^\a\epsilon$)
\begin{align}
{1\over 2}{\bar{\epsilon}}^{i}_{2}\gamma^{a}\epsilon_{1 i}+{1\over 2}{\bar{\epsilon}}_{2 i}\gamma^{a}{\epsilon}^{i}_{1}&=-{i\over 4}\left[ \bar\chi_{2\,+}^i\gamma^a \epsilon_{ij} \chi_{1\,-}^j-\bar\chi_{2\,-}^i\gamma^a \epsilon_{ij} \chi_{1\,+}^j\right] \\
&={i\over 2} \bar\chi_{2}^i\gamma_*\gamma^a \epsilon_{ij} \chi_{1}^j \,.
\end{align}
They are (computation of $\bar\chi_{2}^i\gamma_*\gamma^\mu \epsilon_{ij} \chi_{1}^j\partial_\mu= e_a^{\, \mu}\bar\chi_{2}^i\gamma_*\gamma^a \epsilon_{ij} \chi_{1}^j\partial_\mu$)  
  \begin{align}
0:&\ \    \left[i \cos \tilde\phi\  \bar \chi^i_{0}(\tau_1\otimes\tau_3)  \epsilon_{ij}\chi^j_0+i\sin\tilde\phi\  \bar \chi^i_0(\tau_2\otimes \tau_3) \epsilon_{ij} \chi^j_{0}\right]{\partial_{\tilde\theta}\over \tilde{r}} \\
1:&  \left[i\cot \tilde\theta\cos\tilde \phi\bar\chi^i_{0}(\tau_2\otimes \tau_3)\epsilon_{ij} \chi_{0}^j -i\cot \tilde\theta\sin\tilde \phi\bar\chi_{0}^i(\tau_1\otimes \tau_3)\epsilon_{ij} \chi_{0}^j +i \bar\chi_{0}^i(\tau_3\otimes \tau_3) \epsilon_{ij} \chi_{0}^j\right] {\partial_{\tilde\phi}\over \tilde{r}} \\
2:&\ \  \left[i \cos  \phi\  \bar \chi^i_{0}(1\otimes\tau_2)  \epsilon_{ij}\chi^j_0-i\sin \phi\  \bar \chi^i_0(1\otimes \tau_1) \epsilon_{ij} \chi^j_{0}\right]{\partial_{\theta}\over r}\\
3:&  \left[-i\cot  \theta\cos  \phi\  \bar\chi^i_{0}(1\otimes \tau_1)\epsilon_{ij} \chi_{0}^j -i\cot  \theta\sin  \phi\ \  \bar\chi_{0}^i(1\otimes \tau_2)\epsilon_{ij} \chi_{0}^j +  \bar\chi_{0}^i(1\otimes 1) \epsilon_{ij} \chi_{0}^j  \right]{\partial_{\phi}\over r}
\end{align}
 These are precisely the six Killing vectors of $S^2_{r}\times S^2_{\tilde r}$.\\

\section{BPS equations}
\label{BPSequ} 
The BPS equations associated to the spinor $\chi_{(0)}^i=(\uparrow,\uparrow,\uparrow)+(\downarrow,\downarrow,\downarrow)$ are explicitly
\begin{eqnarray}
\left(1+\gamma_\ast\right)\left[Y_{11}\chi^1_++A\left[\frac{1}{2}X\left(\frac{i}{r}-\frac{1}{\tilde r}\right)\Gamma+i\slashed{D}X+\frac{1}{4}\gamma^{ab}{\cal{F}}_{ab}-i\left[X,\bar X\right]+ Y_{12}\right]\chi^2_+\right]&=&0\nonumber\\
\left(1+\gamma_\ast\right)\left[\left[\frac{1}{2}X\left(\frac{i}{r}-\frac{1}{\tilde r}\right)\Gamma+i\slashed{D}X+\frac{1}{4}\gamma^{ab}{\cal{F}}_{ab}-i\left[X,\bar X\right]- Y_{21}\right]\chi^1_+-AY_{22} \chi^2_+\right]&=&0\nonumber
\end{eqnarray}
and
\begin{eqnarray}
\left(1-\gamma_\ast\right)\left[A Y^{11}\chi^2_+-\left[\frac{1}{2}\bar X\left(\frac{i}{r}-\frac{1}{\tilde r}\right)\Gamma+i\slashed{D}\bar X+\frac{1}{4}\gamma^{ab}{\cal{F}}_{ab}+i\left[X,\bar X\right]+ Y^{12}\right]\chi^1_+\right]&=&0\nonumber\\
\left(1-\gamma_\ast\right)\left[A\left[\frac{1}{2}\bar X\left(\frac{i}{r}-\frac{1}{\tilde r}\right)\Gamma+i\slashed{D}\bar X+\frac{1}{4}\gamma^{ab} {\cal{F}}_{ab}+i\left[X,\bar X\right]- Y^{21}\right]\chi^2_++Y^{22} \chi^1_+\right]&=&0\nonumber
\end{eqnarray}
For the spinor $\chi_{(0)}^i= (\uparrow,\uparrow,\uparrow)+(\downarrow,\downarrow,\downarrow)$ given in appendix (\ref{stereographic}), they read as 
\begin{eqnarray}
{-\bar w\bar z(\left(-i{\cal F}_{\hat{0}\hat{1}}-i{\cal F}_{\hat{2}\hat{3}}\right)-i\left(\frac{i}{r}-\frac{1}{\tilde r}\right)X+2i\left[X,\bar X\right]+2Y_{12})-\left(-i{\cal F}_{\hat{0}\hat{2}}-{\cal F}_{\hat{0}\hat{3}}-{\cal F}_{\hat{1}\hat{2}}+i{\cal F}_{\hat{1}\hat{3}}\right)}&+\nonumber\\
{2\bar z\left(D_{\hat{1}}+iD_{\hat{0}}\right)X+2\bar w\left(D_{\hat{2}}-iD_{\hat{3}}\right)X}-i Y_{11}&=0\nonumber\\
{-\bar w\bar z(\left(i{\cal F}_{\hat{0}\hat{2}}-{\cal F}_{\hat{0}\hat{3}}-{\cal F}_{\hat{1}\hat{2}}-i{\cal F}_{\hat{1}\hat{3}}\right))-\left[-i{\cal F}_{\hat{0}\hat{1}}-i{\cal F}_{\hat{2}\hat{3}}-i\left(\frac{i}{r}-\frac{1}{\tilde r}\right)X-2i\left[X,\bar X\right]-2Y_{12}\right]}&+\nonumber\\
{2\bar w\left(D_{\hat{1}}-iD_{\hat{0}}\right)X-2\bar z\left(D_{\hat{2}}+iD_{\hat{3}}\right)X}+iwzY_{11}&=0\nonumber\\
\left(-i{\cal F}_{\hat{0}\hat{1}}-i{\cal F}_{\hat{2}\hat{3}}\right)-i\left(\frac{i}{r}-\frac{1}{\tilde r}\right)X+2i\left[X,\bar X\right]+2Y_{12}+zw\left(-i{\cal F}_{\hat{0}\hat{2}}-{\cal F}_{\hat{0}\hat{3}}-{\cal F}_{\hat{1}\hat{2}}+i{\cal F}_{\hat{1}\hat{3}}\right)&+\nonumber\\
2w\left(D_{\hat{1}}+iD_{\hat{0}}\right)X+2z\left(D_{\hat{2}}-iD_{\hat{3}}\right)X-{i\bar w\bar z Y_{22}}&=0\nonumber\\
\left(i{\cal F}_{\hat{0}\hat{2}}-{\cal F}_{\hat{0}\hat{3}}-{\cal F}_{\hat{1}\hat{2}}-i{\cal F}_{\hat{1}\hat{3}}\right)+zw\left[-i{\cal F}_{\hat{0}\hat{1}}-i{\cal F}_{\hat{2}\hat{3}}-i\left(\frac{i}{r}-\frac{1}{\tilde r}\right)X-2i\left[X,\bar X\right]-2Y_{12}\right]&+\nonumber\\
2z\left(D_{\hat{1}}-iD_{\hat{0}}\right)X-2w\left(D_{\hat{2}}+iD_{\hat{3}}\right)X+{iY_{22}}&=0\nonumber
\end{eqnarray} 
and
\begin{eqnarray}
w\left(-i{\cal F}_{\hat{0}\hat{2}}+{\cal F}_{\hat{0}\hat{3}}-{\cal F}_{\hat{1}\hat{2}}-i{\cal F}_{\hat{1}\hat{3}}\right)+z\left[i{\cal F}_{\hat{0}\hat{1}}-i{\cal F}_{\hat{2}\hat{3}}+i\left(\frac{i}{r}-\frac{1}{\tilde r}\right)\bar X+2i\left[X,\bar X\right]-2Y_{12}\right]&-\nonumber\\
2wz\left(D_{\hat{1}}+iD_{\hat{0}}\right)\bar X+2\left(D_{\hat{2}}+iD_{\hat{3}}\right)\bar X-{i\bar w Y_{11}}&=0\nonumber\\
w\left[-i{\cal F}_{\hat{0}\hat{1}}+i{\cal F}_{\hat{2}\hat{3}}-i\left(\frac{i}{r}-\frac{1}{\tilde r}\right)\bar X+2i\left[X,\bar X\right]-2Y_{12}\right]+z\left(-i{\cal F}_{\hat{0}\hat{2}}-{\cal F}_{\hat{0}\hat{3}}+{\cal F}_{\hat{1}\hat{2}}-i{\cal F}_{\hat{1}\hat{3}}\right)&+\nonumber\\
2\left(D_{\hat{1}}-iD_{\hat{0}}\right)\bar X+2wz\left(D_{\hat{2}}-iD_{\hat{3}}\right)\bar X-{i\bar zY_{11}}&=0\nonumber\\
{\bar z\left(-i{\cal F}_{\hat{0}\hat{2}}+{\cal F}_{\hat{0}\hat{3}}-{\cal F}_{\hat{1}\hat{2}}-i{\cal F}_{\hat{1}\hat{3}}\right)+\bar w\left[i{\cal F}_{\hat{0}\hat{1}}-i{\cal F}_{\hat{2}\hat{3}}+i\left(\frac{i}{r}-\frac{1}{\tilde r}\right)\bar X+2i\left[X,\bar X\right]-2Y_{12}\right]}&+\nonumber\\
{2\left(D_{\hat{1}}+iD_{\hat{0}}\right)\bar X-2\bar w \bar z\left(D_{\hat{2}}+iD_{\hat{3}}\right)\bar X}+iz Y_{22}&=0\nonumber\\
{\bar z\left[-i{\cal F}_{\hat{0}\hat{1}}+i{\cal F}_{\hat{2}\hat{3}}-i\left(\frac{i}{r}-\frac{1}{\tilde r}\right)\bar X+2i\left[X,\bar X\right]-2Y_{12}\right]+\bar w\left(-i{\cal F}_{\hat{0}\hat{2}}-{\cal F}_{\hat{0}\hat{3}}+{\cal F}_{\hat{1}\hat{2}}-i{\cal F}_{\hat{1}\hat{3}}\right)}&-\nonumber\\
{2\bar w\bar z\left(D_{\hat{1}}-iD_{\hat{0}}\right)\bar X-2\left(D_{\hat{2}}-iD_{\hat{3}}\right)\bar X}+iwY_{22}&=0\nonumber
\end{eqnarray}
   
The non-vanishing field configurations are \rf{susyconfg}
   \beq
     F_{\hat 0\hat 1}={2\over \tilde r} \text{Re}(X)={\tilde B \over 2\tilde r^2} \qquad  F_{\hat 2\hat 3}={2\over  r} \text{Im}(X)={B\over 2 r^2}\,.
  \eeq

 \vfill\eject
 
\bibliography{refs}

\providecommand{\href}[2]{#2}\begingroup\raggedright\begin{thebibliography}{10}

\bibitem{Pestun:2007rz}
V.~Pestun, ``{Localization of gauge theory on a four-sphere and supersymmetric
  Wilson loops},'' {\em Commun.Math.Phys.} {\bf 313} (2012) 71--129,
\href{http://www.arXiv.org/abs/0712.2824}{{\tt 0712.2824}}.

\bibitem{Gerchkovitz:2014gta}
E.~Gerchkovitz, J.~Gomis, and Z.~Komargodski, ``{Sphere Partition Functions and
  the Zamolodchikov Metric},''
\href{http://www.arXiv.org/abs/1405.7271}{{\tt 1405.7271}}.

\bibitem{Gomis:2014woa}
J.~Gomis and N.~Ishtiaque, ``{Kahler Potential and Ambiguities in 4d N=2
  SCFTs},''
\href{http://www.arXiv.org/abs/1409.5325}{{\tt 1409.5325}}.

\bibitem{Hama:2012bg}
N.~Hama and K.~Hosomichi, ``{Seiberg-Witten Theories on Ellipsoids},'' {\em
  JHEP} {\bf 1209} (2012) 033,
\href{http://www.arXiv.org/abs/1206.6359}{{\tt 1206.6359}}.

\bibitem{Klare:2013dka}
C.~Klare and A.~Zaffaroni, ``{Extended Supersymmetry on Curved Spaces},'' {\em
  JHEP} {\bf 1310} (2013) 218,
\href{http://www.arXiv.org/abs/1308.1102}{{\tt 1308.1102}}.

\bibitem{Bawane:2014uka}
A.~Bawane, G.~Bonelli, M.~Ronzani, and A.~Tanzini, ``{$\mathcal{N}=2$
  supersymmetric gauge theories on $S^2\times S^2$ and Liouville Gravity},''
\href{http://www.arXiv.org/abs/1411.2762}{{\tt 1411.2762}}.

\bibitem{Benini:2012ui}
F.~Benini and S.~Cremonesi, ``{Partition functions of $N=(2,2)$ gauge theories
  on $S^2$ and vortices},''
\href{http://www.arXiv.org/abs/1206.2356}{{\tt 1206.2356}}.

\bibitem{Doroud:2012xw}
N.~Doroud, J.~Gomis, B.~Le~Floch, and S.~Lee, ``{Exact Results in D=2
  Supersymmetric Gauge Theories},''
\href{http://www.arXiv.org/abs/1206.2606}{{\tt 1206.2606}}.

\bibitem{Festuccia:2011ws}
G.~Festuccia and N.~Seiberg, ``{Rigid Supersymmetric Theories in Curved
  Superspace},'' {\em JHEP} {\bf 1106} (2011) 114,
\href{http://www.arXiv.org/abs/1105.0689}{{\tt 1105.0689}}.

\bibitem{deWit:1980tn}
B.~de~Wit, J.~van Holten, and A.~Van~Proeyen, ``{Structure of N=2
  Supergravity},'' {\em Nucl.Phys.} {\bf B184} (1981)
77.

\bibitem{freedman12}
D.~Z. Freedman and A.~V. Proeyen, {\em Supergravity}.
\newblock Cambridge University Press, New York, USA, 2012.

\bibitem{deWit:1979ug}
B.~de~Wit, J.~van Holten, and A.~Van~Proeyen, ``{Transformation Rules of N=2
  Supergravity Multiplets},'' {\em Nucl.Phys.} {\bf B167} (1980)
186.

\bibitem{deWit:1979pq}
B.~de~Wit and J.~van Holten, ``{Multiplets of Linearized SO(2) Supergravity},''
  {\em Nucl.Phys.} {\bf B155} (1979)
530.

\bibitem{deWit:1982na}
B.~de~Wit, R.~Philippe, and A.~Van~Proeyen, ``{The Improved Tensor Multiplet in
  $N=2$ Supergravity},'' {\em Nucl.Phys.} {\bf B219} (1983)
143.

\bibitem{parisLect}
A.~V. Proeyen, ``N = 2 supergravity in d = 4, 5, 6 and its matter couplings.''
  \url{http://itf.fys.kuleuven.be/~toine/LectParis.pdf}.

\bibitem{Nekrasov:2002qd}
N.~A. Nekrasov, ``{Seiberg-Witten prepotential from instanton counting},'' {\em
  Adv. Theor. Math. Phys.} {\bf 7} (2004) 831--864,
\href{http://www.arXiv.org/abs/hep-th/0206161}{{\tt hep-th/0206161}}.

\end{thebibliography}\endgroup
\end{document}